\documentclass[twocolumn,showpacs,prl]{revtex4}

\begin{document}

\newcommand{\ket}[1]{\ensuremath{|#1 \rangle}}
\newcommand{\bra}[1]{\ensuremath{\langle #1|}}
\newcommand{\braket}[2]{\ensuremath{\langle #1|#2 \rangle}}
\newcommand{\ketbra}[2]{\ensuremath{|#1 \rangle \langle #2|}}
\newcommand{\ro}[1]{\ensuremath{|#1 \rangle \langle #1|}}
\newcommand{\av}[1]{\ensuremath{\langle #1 \rangle}}

\newcommand{\real}{\ensuremath{\mathrm{Re}}}
\newcommand{\trace}{\ensuremath{\textsf{Tr}}}

\newcommand{\id}{\ensuremath{\mathsf{1}}}
\newcommand{\iden}{\ensuremath{\openone}}
\newcommand{\R}{\ensuremath{{\sf R\hspace*{-0.9ex}\rule{0.15ex}
{1.5ex}\hspace*{0.9ex}}}}
\newcommand{\N}{\ensuremath{{\sf N\hspace*{-1.0ex}\rule{0.15ex}
{1.3ex}\hspace*{1.0ex}}}}
\newcommand{\Q}{\ensuremath{{\sf Q\hspace*{-1.1ex}\rule{0.15ex}
{1.5ex}\hspace*{1.1ex}}}}
\newcommand{\C}{\ensuremath{{\sf C\hspace*{-0.9ex}\rule{0.15ex}
{1.3ex}\hspace*{0.9ex}}}}

\newcommand{\h}[1]{\ensuremath{\mathcal{H}_{#1}}}

\newcommand{\me}{\ensuremath{\mathrm{e}}}
\newcommand{\mi}{\ensuremath{\mathrm{i}}}

\newcommand{\de}{\ensuremath{\mathrm{d}}}
\newcommand{\dd}[2]{\ensuremath{\frac{\mathrm{d}#1}{\mathrm{d}#2}}}
\newcommand{\ddd}[2]{\ensuremath{\frac{\mathrm{d}^2#1}{\mathrm{d}#2^2}}}

\newcommand{\ot}[2]{\ensuremath{\left( \begin{array}{c} #1 \\ #2
\end{array} \right)}}
\newcommand{\oth}[3]{\ensuremath{\left( \begin{array}{c} #1 \\ #2 \\ #3
\end{array} \right)}}
\newcommand{\twtw}[4]{\ensuremath{\left( \begin{array}{cc} #1 & #2 \\
#3 & #4 \end{array} \right)}}
\newcommand{\thth}[9]{\ensuremath{\left( \begin{array}{ccc} #1 & #2 & #3
\\ #4 & #5 & #6 \\ #7 & #8 & #9 \end{array} \right)}}

\newcommand{\expp}[1]{\ensuremath{\me^{\mi\hat{H}#1}}}
\newcommand{\expm}[1]{\ensuremath{\me^{-\mi\hat{H}#1}}}

\newcommand{\q}{\ensuremath{q_{ai}}}

\title{Entanglement without nonlocality}
\author{C. Hewitt-Horsman$^{\dagger}$, V. Vedral$^{\diamond}$}
\address{$^{\dagger}$Department of Physics and Astronomy, University College, London WC1E 6BT, United Kingdom\\
$^{\diamond}$School of Physics and Astronomy, University of Leeds, Leeds LS2 9JT, United Kingdom}

\begin{abstract}
We consider the characterization of entanglement from the perspective of a Heisenberg formalism. We derive an original two-party generalized separability criteria, and from this describe a novel physical understanding of entanglement. We find that entanglement may be considered as fundamentally a local effect, and therefore as a separable computational resource from nonlocality. We show how entanglement differs from correlation physically, and explore the implications of this new conception of entanglement for the notion of classicality. We find that this understanding of entanglement extends naturally to multipartite cases.
\end{abstract}

\maketitle

\section{Introduction}

Entanglement is a key concept in quantum mechanics, and plays a central role in the field of quantum computation. It acts as a resource that enables us to perform both faster calculations \cite{shor} and otherwise impossible communication protocols \cite{teleport, nandc}. It is also a concept of considerable importance in the foundations of quantum theory. Entanglement remains, however, notoriously difficult to understand, and the tasks of qualifying and quantifying it are extremely challenging. In this paper we will present a new physical understanding of the concept of entanglement. We will look at how and why systems become entangled, and what it means in conceptual terms for systems to be either entangled or separable. While the formalisation of entanglement will be central to our investigation, we will not primarily be concerned with giving new calculational tools for the recognition of separability or the computation of entanglement quantities. Rather, we are trying to understand entanglement as a physical property of systems. This understanding is to be offered in the hope that it will be part of a full picture of this important, yet elusive, concept. In particular, we will find that the picture we gain of entanglement may be of use in understanding multipartite entanglement, and the relation of entangled subsystems to a larger entangled system. At present, these are both very poorly understood

Entanglement is often considered to be a fundamentally \emph{nonlocal} effect, and indeed as the generator of nonlocality in quantum mechanics (see for example  \cite{martinreview}). Such a view finds expression in the standard description of entangled systems, that shared entanglement enables a system to influence another instantaneously, no matter how far apart they are (whilst, of course, preserving the same local statistics -- an effect know as \emph{statistical locality} \cite{statlocl}, arising from the well-known no-signalling theorem \cite{nosig}). This is given as the mechanism behind effects such as teleportation, where quantum information appears to `jump' between separated qubits. Entanglement also played a key role in the first explicit discussion of nonlocality in quantum mechanics, the famous paper by Einstein, Podolsky and Rosen \cite{epr}. The connection between entanglement and nonlocality is, however, a conceptual one, rather than a theorem of quantum mechanics. The exact nature of their relationship has not been established, and this will be considered in detail in this paper. We will find that, contrary to the standard belief, entanglement can best be understood as a purely local effect. 

In order to make our investigation of entanglement as general as possible, we are going to use a no-collapse regime. This is of particular relevance in quantum computation, where coherent entangled states are generally used throughout the algorithm \cite{qft}. We will also be using a Heisenberg, rather than a Schr\"{o}dinger, representation in which to frame our results. There has recently been interest in the Heisenberg approach to analyse computation \cite{dh, chris} and spin chains \cite{spin}, where entanglement plays a vital role \cite{spinchainent1,spinchainent2,spinchainent3}. A Heisenberg-picture specification of entanglement therefore would also have specific computational use in these analyses. Furthermore, the Heisenberg approach of emphasising operators rather than states accords closely with the approach to many-body systems in solid-state physics. Entanglement is particularly important in these systems as correlations will influence the properties of materials \cite{vlatko}. The understanding of entanglement given here will therefore find many uses in such projects. We will use a particular type of Heisenberg formalism, that of the Deutsch-Hayden approach. The primary reason for using this is its transparency to the localisation of qubits. When we are dealing with questions of locality it is very important that we are able to say that the (non)locality that we have found is real, rather than an artefact of a particular formalism. In the standard Schr\"{o}dinger representation a joint system is in general given its fullest description by giving a joint characterisation (a joint density matrix). Individual density matrices in general do not contain the full amount of data. In the Deutsch-Hayden approach, however, each system has its own descriptor, which is necessarily localized when the system is localized. We therefore do not have extraneous non-local descriptions of joint systems. 

We will therefore start in the first section by outlining the formalism that we will use. In the next section we will then discuss the understanding of correlation and pure state entanglement that we have found using this approach. After this, we will begin to generalize this to mixed states, and will derive an entirely new necessary and sufficient separability criterion for two mixed qubits. The physical explication of this will take place in the next section, giving us the physical description of entanglement that we have been looking for, and expanding on the implications that this understanding of entanglement has for, amongst other areas, quantum information theory. In the final section we will make good our claim that entanglement may be constructed as a purely local element of quantum mechanics.


\section{The Deutsch-Hayden Heisenberg approach}

The Deutsch-Hayden formalism was presented in \cite{dh} and expanded in \cite{mev}. It is based on the development of stabilizer theory by Gottesmann \cite{gottesman, gott2}, and is closely related to it. In this approach we start with a standard universal Heisenberg state, by convention usually fixed as $\ket{\mathbf{)}}$, and time-evolving operators. The evolution of a given system, $a$, is represented by tracking its \emph{descriptors}, given by 
$$\mathbf{q}_a(t) = U^\dagger \mathbf{q}_a(0) U$$
\noindent These are chosen as they form a basis in the Hilbert-Schmidt space of operators for the system, so can be used to reconstruct the time-evolved form of any operator (and hence make physical predictions). In the terms of stabilizer theory, they are the individual elements of the stabilizer group. By convention, we choose
\begin{equation}\mathbf{q}_a(0) = \iden^{\otimes a-1} \otimes \mathbf{\sigma} \otimes \iden^{\otimes n-a} \label{one}\end{equation}
\noindent for an n-qubit array. The descriptors for a joint system are the simple combination of descriptors for individual systems, \emph{eg} $q_{12p} = q_{1i}q_{2j}$. The most natural way to move between this formalism and the Schr\"{o}dinger representation is through the density matrix. In terms of the descriptors, for two qubits this takes the simple form
$$\rho_{12} = \sum_{i,j=0}^3  \av{q_{1i}q_{2j}} \sigma_i \otimes \sigma_j$$
If we look at equation (\ref{one}) we can see can see that, after a time, in general the descriptor for any given qubit will span the space of all qubits in the system. However, it is often possible to reduce the dimensionality of the descriptor for all practical purposes. Such \emph{reduced descriptors} are formed by ignoring all components not on the specified spaces. For example, the reduced descriptor of $\mathbf{q}_a$ on space $N$ is written
$$ [ \mathbf{q}_a ]_N = \mathbf{q}_a \in \h{N}$$
As a concrete example, take the descriptor component $q_{1x} = \sigma_z \otimes \openone \otimes \sigma_x$, which spans $\h{123}$. The reduced descriptor on $\h{1}$ only is $[ q_{1x} ]_{1} = \sigma_z$.
In order for us to be able to substitute a reduced descriptor for a full one, we need to make sure that the reduced version gives us the same results as the full one. In our above example we can see that this is not the case: taking a look at the average values of the operators, we see that
$$ \av{q_{1x}} \neq \av{ [ q_{1x} ]_{12}}$$
In fact, it is only in very specific circumstances that the reduced descriptors may be used \cite{mev}. In order to reduce a descriptor to the space of a given system, that system must be pure. In our example, qubit 1 is not in a pure state as we cannot reduce its descriptor to the space $\h{1}$ whilst preserving its predictions. It is possible, however, that taken together with qubit 3 it is in a pure state, that qubit 3 purifies qubit 1 -- but we would need to see the remaining descriptors for all three systems in order to work that out. If this were the case, then we would be able to write the $x$-component of the reduced descriptor for qubit 1 as
$$ [ q_{1x} ]_{13} = \sigma_z \otimes \sigma_x$$
We may, hoever, want to write a descriptor on the space of a system that is not pure. This is done by representing it by a convex sum of descriptors that are pure on that space \cite{mev}. For example, if we have two qubits which are both mixed, and we denote by $\mathbf{q}_1^i$ all the possible descriptors on $\h{12}$ of a qubit that is pure on $\h{1}$, then we can write
$$ \av{\mathbf{q}_1} = \av{\sum_i w_i [ \mathbf{q}_1^i ]_1}$$
\noindent In other words, we can use $\sum_i w_i [ \mathbf{q}_1^i ]_1$ as the reduced descriptor on $\h{1}$.

\section{Correlation and the reduced descriptors}

Using this formalism, we have two conditions for pure state separability. The first comes from considering the average values of operators in separable states. If $A$ and $B$ are operators then
\begin{equation} \av{AB} = \av{A}\av{B} \ \ \ \Rightarrow \av{\mathbf{q}_1\mathbf{q}_2} = \av{\mathbf{q}_1}\av{\mathbf{q}_2}\label{msete}\end{equation}
\noindent The second is the use of the reduced descriptors:
\begin{equation} \av{\mathbf{q}_1\mathbf{q}_2} = \av{ [\mathbf{q}_1]_1 \ [\mathbf{q}_2]_2}  \label{mseti}\end{equation}
\noindent These are not, however, entirely separate conditions (nor would we expect them to be) as (\ref{mseti}) is derived from (\ref{msete}) \cite{mev}.

The ability to use the reduced descriptors in separable states points to one element of an understanding entanglement for pure states. Consider $\mathbf{q}_1 \in \h{12}$ where $\av{\mathbf{q}_1} = \av{[\mathbf{q}_1]_{1}}$. The neglected element $[\mathbf{q}_1]_{2}$ represents the evolution of the first system on the space of the second. That is, it is a record of the past interaction of the first system with the second. If the systems are not entangled then this past interaction becomes irrelevant for the description of the first system -- it is as if they had never interacted. So, if systems \emph{are} entangled then we cannot ignore one if we are looking at the past evolution of another; it is irreducibly part of the history of that system.

This is not, however, the most general form of entanglement. For this, we need to look at entanglement between systems in mixed states. We have, of course, a different criterion for separability in this situation:
$$\rho_{12} = \sum_i w_i \rho_{1i}\otimes\rho_{2i}$$
Equation (\ref{msete}) now becomes the criteria for systems to be uncorrelated rather than separable. However, the relation of (\ref{mseti}) to correlation is more complicated, and will now investigate it. 

A useful system for these purposes is a 3-qubit Greenberger-Horne-Zeilinger (GHZ) state. In Schr\"{o}dinger notation this is written (unnormalized) as $\ket{000} + \ket{111}$. This state has the property that it is overall pure and entangled, but if any one qubit is traced out then the remaining pair is in a separable mixed state. We can (not, of course, uniquely) construct a GHZ state from a singlet pair and a ground state system by the operation $CNOT(1,3)(\ket{00} + \ket{11})\ket{0}$. In terms of descriptors this gives us
\begin{eqnarray*}\mathbf{q}_1 & = & \left( \sigma_z \otimes \sigma_x \otimes \sigma_x, \  -\sigma_y \otimes \sigma_x \otimes \sigma_x ,\  \sigma_x \otimes \iden \otimes \iden \right)\\
\mathbf{q}_2 & = &  \left(  \iden \otimes \sigma_x \otimes \iden ,\ \sigma_x \otimes \sigma_y \otimes \iden ,\ \sigma_x \otimes \sigma_z \otimes \iden \right)\\
\mathbf{q}_3 & = & \left(  \iden \otimes \iden \otimes \sigma_x ,\ \sigma_x \otimes \iden \otimes \sigma_y ,\ \sigma_x \otimes \iden \otimes \sigma_z  \right)          \end{eqnarray*}             
We can see immediately that each pair is mixed, as the reduced descriptors do not give the same average values as the full descriptors. For example, if we take system 1 and 3, thus neglecting $\h{2}$, we are left with
\begin{eqnarray*}[\mathbf{q}_1]_{13} &  = &   \left(  \sigma_z \otimes \sigma_x ,\ -\sigma_y  \otimes \sigma_x ,\ \sigma_x \otimes \iden \right) \\
{[\mathbf{q}_3]_{13}} &  =  & \left(  \iden  \otimes \sigma_x ,\ \sigma_x \otimes  \sigma_y ,\ \sigma_x \otimes  \sigma_z  \right) \end{eqnarray*}
\noindent which gives us $\av{[\mathbf{q}_1\mathbf{q}_3]_{13}} \neq \av{\mathbf{q}_1\mathbf{q}_3}$.
We also see that each pair is correlated, through its $q_zq_z$ component -- \emph{ie} for each pair $\av{q_{az}q_{bz}} \neq \av{q_{az}}\av{q_{bz}}$. The obvious thing to assume at this point is that the relationship between correlation and the use of reduced descriptors holds, in that if systems $a$ and $b$ are correlated then it is not possible to neglect the components of \emph{eg} $\mathbf{q}_a$ on $\h{b}$. The GHZ descriptors, however, tell us that this is an incorrect assumption. Take systems 2 and 3. We see that they are correlated, and yet we have $ \av{\mathbf{q}_2\mathbf{q}_3} = \av{[\mathbf{q}_2]_{12}[\mathbf{q}_3]_{13}}$, thus breaking the link between correlation and use of the reduced descriptors. Note that the product $\av{[\mathbf{q}_2]_{12}[\mathbf{q}_3]_{13}}$ is not straightforwardly a tensor product as it is in the pure-state case, as here we have a non-trivial vector product on \h{1}. To make this product with the reduced descriptors, the original descriptors are taken and the components to be neglected are then replaced by $\iden$. The product is then between two objects on \h{123}.

What is going on here can be seen from the combined descriptors: it is the components on $\h{1}$ which give the average values of the combinations different from the combined averages. That is, systems 2 and 3 are correlated by their (separate) past interactions with system 1. What the reduced descriptors are telling us here is not the correlation between the systems but the systems' past interactions. Consider again how we constructed this particular set of GHZ descriptors. The first two systems started off in a singlet state, unentangled and uncorrelated with the third system which was in the zero state. System 1 then interacted with first system 2 and then with system 3 -- at no time did systems 2 and 3 interact with each other. It is this lack of past interaction that is being picked up by the fact that the reduced descriptors are sufficient fully to describe the physically measurable properties of the systems.

The next obvious assumption to make is that as the use of the reduced descriptors is not telling us about correlation in the mixed case, perhaps it will tell us about \emph{entanglement}? This would then be identical to the pure case, and we could use our understanding of entanglement as needing to take past interaction with other systems into account when looking at present measurable properties. Unfortunately this is easily shown not to be the case. If we take the reduced descriptors for the pairs (12) and (13) then they do not give the same average values as the full descriptors.

If we look at pairs (12) and (23) however, we find that the reduced descriptors do not preserve the average values of the full descriptors, and it is interesting to look at which values differ. Let us look at the reduced descriptors in full. Writing the reduced descriptors all on $\h{123}$ we have for the pair (12):
\begin{eqnarray*} [\mathbf{q}_1]_{13} & = & \left( \sigma_z \otimes \iden \otimes \sigma_x ,\ -\sigma_y \otimes \iden \otimes \sigma_x ,\ \sigma_x \otimes \iden \otimes \iden \right)  \\
{[\mathbf{q}_2]_{23}} & = & \left(  \iden \otimes \sigma_x \otimes \iden ,\ \iden \otimes \sigma_y \otimes \iden ,\ \iden \otimes \sigma_z \otimes \iden \right) \end{eqnarray*}
 \noindent And for pair (13):
\begin{eqnarray*}[\mathbf{q}_1]_{12} & =&  \left( \sigma_z \otimes \sigma_x \otimes \iden ,\ -\sigma_y \otimes \sigma_x \otimes \iden ,\ \sigma_x \otimes \iden \otimes \iden  \right) \\ 
{[\mathbf{q}_3]_{23}} & =  & \left( \iden \otimes \iden \otimes \sigma_x ,\ \iden \otimes \iden \otimes \sigma_y ,\ \iden \otimes \iden \otimes \sigma_z  \right)\end{eqnarray*}                  
If we compare these with the full descriptors then we see that for each pair all of the descriptors give the same average values \emph{except} the components $q_{1z}q_{2z}$ and $q_{1z}q_{3z}$. For these products the reduced descriptors give 0 rather than 1. What is interesting is that, as we saw above, these are the components which show the correlations between the systems. With the reduced descriptors, all the average values are the same \emph{except} for those for the components which show the correlation. This lets us see that the correlation has arisen in the case of these two pairs from the interaction of each member of the pair with the other, rather than with a third party. If you take away this history of this mutual interaction (by using the reduced descriptors) then you lose the correlation. 

Thus we have retained a partial connection between the reduced descriptors and correlation. If the reduced descriptors retain the correlation then we know that it has been brought about by past interaction with a third party. However, if the correlation is no longer there with the reduced descriptors then we can see that it has been produced directly from past interaction between the two systems alone.

\section{Generalized separability criteria for descriptors}

We turn our attention now to the subject of mixed state entanglement. In terms of the descriptors, the standard mixed-state separability criterion is $\av{\mathbf{q}_1\mathbf{q}_2} = \sum_i w_i \av{\mathbf{q}_{1i}} \av{\mathbf{q}_{2i}}$. As it stands, this is rather like (\ref{msete}): it is not obvious either where it comes from in the Deutsch-Hayden approach, or what it means physically in that picture. We would therefore like to find a mixed-state analogue of (\ref{mseti}). Let us now look at the descriptors written on only a limited number of spaces. Now, any individual descriptor may be given on any Hilbert space by a weighted sum of basis descriptors for that space \cite{mev}. Therefore the ability to write a given descriptor without reference to the Hilbert space of another system tells us nothing about whether the two systems are entangled or not. But what if we consider the ability to write the overall descriptors for the combined system that we are looking at as a simple combination of such descriptors for the individual systems that do not have support on the Hilbert space of the other system?

In formal terms what we have is this. We can always represent the individual $\mathbf{q}_1$ and $\mathbf{q}_2$ as
\begin{eqnarray} \av{\mathbf{q}_1} & = & \sum_i \mu_i \trace ( \ketbra{00}{00} U_i^\dagger \mathbf{\sigma} \otimes \iden U_i ) \label{hip}\\
\av{\mathbf{q}_2} & = & \sum_j \nu_j \trace ( \ketbra{00}{00}  U_j^{\prime\dagger} \mathbf{\sigma} \otimes \iden U_j^\prime ) \label{hop}\end{eqnarray}
\noindent where $U_i$ acts only on $\h{13}$ and $U_j^\prime$ only on $\h{23}$. However, in general when we want the descriptors for the overall system (writing on $\h{132}$ for convenience),
$$ \av{\mathbf{q}_1 \mathbf{q}_2} \neq \sum_{ij} \lambda_{ij} \trace ( \ketbra{000}{000}  \ U_i^\dagger \mathbf{\sigma} \otimes \iden U_i \otimes \iden \ \ \iden \otimes U_j^{\prime \dagger} \iden \otimes \mathbf{\sigma} U_j^\prime )$$
\noindent What we are suggesting here is that there \emph{is} an equality, if and only if the two systems are separable. In other words, that
\begin{eqnarray}\lefteqn{\sum_{ij} \lambda_{ij} \trace ( \ketbra{000}{000}  \ U_i^\dagger \mathbf{\sigma} \otimes \iden U_i \otimes \iden \ \ \iden \otimes U_j^{\prime \dagger} \iden \otimes \mathbf{\sigma} U_j^\prime )}\nonumber\\
& & = \sum_i w_i \trace ( \ketbra{00}{00}  U_i^\dagger \mathbf{\sigma} \otimes \iden U_i ) \trace ( \ketbra{00}{00}  U_i^{\prime \dagger}\mathbf{\sigma} \otimes \iden U_i^\prime ) \nonumber\\
& & {} \label{hio}
\end{eqnarray}
\noindent and hence that the condition
$$ \av{\mathbf{q}_1 \mathbf{q}_2} = \sum_{ij} \lambda_{ij} \trace ( \ketbra{000}{000}  \ U_i^\dagger \mathbf{\sigma} \otimes \iden U_i \otimes \iden \ \ \iden \otimes U_j^{\prime \dagger} \iden \otimes \mathbf{\sigma} U_j^\prime )$$
\noindent is a necessary and sufficient separability condition. We shall now prove that this is indeed the case.

Let us look at the left-hand side of (\ref{hio}). Part of the proposition here is that the weights for the different components in the sum, $\lambda_{ij}$ are the products of the weights $\mu_i$ and $\nu_j$ in (\ref{hip}) and (\ref{hop}). That is, we have
\begin{eqnarray*} \lambda_{ij} & = & \mu_i \nu_j\\
{} & = &  \trace ( U_i^\dagger \mathbf{\sigma} \otimes \iden U_i \ U_{13}^\dagger \mathbf{\sigma} \otimes \iden U_{13}) \trace ( U_2^\dagger \ketbra{0}{0} U_2 ) \\
{} & {} & .  \ \trace ( U_j^{\prime\dagger} \mathbf{\sigma} \otimes \iden U_j^\prime \ U_{23}^\dagger \mathbf{\sigma} \otimes \iden U_{23}) \trace ( U_1^\dagger \ketbra{0}{0} U_1 )
\end{eqnarray*}
\noindent At this point we note that $\trace ( U_2^\dagger \ketbra{0}{0} U_2 ) . \trace ( U_1^\dagger \ketbra{0}{0} U_1 )$ is a constant for all components in the sum. For present purposes we can therefore ignore it as part of the normalisation of the descriptor. We can therefore write the left-hand side of (\ref{hio}) as
\begin{eqnarray} \lefteqn{\sum_{ij} \trace ( U_i^\dagger \mathbf{\sigma} \otimes \iden U_i \ U_{13}^\dagger \mathbf{\sigma} \otimes \iden U_{13}) }\nonumber\\
& & {}. \trace ( U_j^{\prime\dagger} \mathbf{\sigma} \otimes \iden U_j^\prime \ U_{23}^\dagger \mathbf{\sigma} \otimes \iden U_{23})  \nonumber\\
& & {}. \trace ( \ketbra{000}{000}  \ U_i^\dagger \mathbf{\sigma} \otimes \iden U_i \otimes \iden \ \ \iden \otimes U_j^{\prime \dagger} \iden \otimes \mathbf{\sigma} U_j^\prime )\label{pfi}\end{eqnarray}
\noindent We will now introduce some notation to make this rather unwieldy expression a little more tractable. Let us define
$$\mathbf{a}_i \otimes \mathbf{b}_i  \equiv  U_i^\dagger \mathbf{\sigma} \otimes \iden U_i \  ; \ \ \mathbf{m}_j \otimes \mathbf{n}_j  \equiv  U_j^{\prime \dagger} \mathbf{\sigma} \otimes \iden U_j^\prime$$

\noindent This gives us
\begin{eqnarray} \lefteqn{\sum_{ij} \trace ( \mathbf{a}_i \otimes \mathbf{b}_i \ U_{13}^\dagger \mathbf{\sigma} \otimes \iden U_{13})  \trace ( \mathbf{m}_j \otimes \mathbf{n}_j  \ U_{23}^\dagger \mathbf{\sigma} \otimes \iden U_{23})}\nonumber \\
& & \hspace{1cm}.\trace ( \ketbra{0}{0} \mathbf{a}_i ) \trace ( \ketbra{0}{0} \mathbf{m}_j ) \trace ( \ketbra{0}{0} ( \mathbf{b}_i \mathbf{n}_j )) \label{tins}
\end{eqnarray}
\noindent In order to look at the behaviour of this double sum, let us now fix the $i$ index, and see what happens when only the $j$ index is varied. For each value of $i$ the variable sum is then
\begin{equation} \sum_j \trace ( \mathbf{m}_j \otimes \mathbf{n}_j  \ U_{23}^\dagger \mathbf{\sigma} \otimes \iden U_{23}) \trace ( \ketbra{0}{0} \mathbf{m}_j ) \trace ( \ketbra{0}{0} ( \mathbf{b}_i \mathbf{n}_j )) \label{pin}\end{equation}
\noindent We now note that in general 
\begin{equation}\trace ( \ketbra{0}{0} ( \mathbf{b}_i \mathbf{n}_j )) \neq \trace ( \ketbra{0}{0} \mathbf{b}_i ) \trace ( \ketbra{0}{0} \mathbf{n}_j ) \label{mel}\end{equation}
\noindent We will now show that when this is the case, the expression (\ref{pin}) sums to zero. In order to show this we will show that for each $j=p$, such that (\ref{mel}) holds, there is a $j=q$ such that
\begin{eqnarray*} \lefteqn{\trace ( \mathbf{m}_p \otimes \mathbf{n}_p  \ U_{23}^\dagger \mathbf{\sigma} \otimes \iden U_{23}) \trace ( \ketbra{0}{0} \mathbf{m}_p ) \trace ( \ketbra{0}{0} ( \mathbf{b}_i \mathbf{n}_p ))} \\
& \hspace{3cm}= & {}  -  \trace ( \mathbf{m}_q \otimes \mathbf{n}_q  \ U_{23}^\dagger \mathbf{\sigma} \otimes \iden U_{23})\\
& \hspace{3cm}& {}. \trace ( \ketbra{0}{0} \mathbf{m}_q ) \trace ( \ketbra{0}{0} ( \mathbf{b}_i \mathbf{n}_q )) \end{eqnarray*}
\noindent To start, we re-write
\begin{eqnarray*} \trace ( \mathbf{m}_j \otimes \mathbf{n}_j  \ U_{23}^\dagger \mathbf{\sigma} \otimes \iden U_{23}) & = & \trace ( U_{23}^\dagger \mathbf{m}_j \otimes \mathbf{n}_j  U_{23}  \mathbf{\sigma} \otimes \iden ) \\
{} & = & \trace (\mathbf{m}_j(t) \otimes \mathbf{n}_j(t) \mathbf{\sigma} \otimes \iden )\\
{} & = & \trace (\mathbf{m}_j(t) \mathbf{\sigma})\trace (\mathbf{n}_j(t) ) 
\end{eqnarray*}
\noindent Let us now look at $U_N$, which we will define as taking us from the $q$-descriptor to the $p$-descriptor:
$$\mathbf{m}_p \otimes \mathbf{n}_p \equiv U_N^\dagger \mathbf{m}_q \otimes \mathbf{n}_q U_N$$
\noindent We will look at the subset of possible $U_N$ such that $U_N = \iden \otimes U_n$. This has the immediate effect that $\mathbf{m}_p = \mathbf{m}_q$ and hence that
\begin{equation}\trace (\mathbf{m}_q(t) \mathbf{\sigma}) = \trace (\mathbf{m}_p(t) \mathbf{\sigma})\label{mptq}\end{equation}
\noindent and that
\begin{equation} \trace ( \ketbra{0}{0} \mathbf{m}_q )  =    \trace ( \ketbra{0}{0} \mathbf{m}_p )  \label{tothmptq}\end{equation}
\noindent We now make use of the fact that a unitary transformation does not change the trace of a vector. As the transformation of $U_N$ is confined to $\h{3}$ then it does not change the trace on that space. In other words, $ \trace ( \mathbf{n}_q ) = \trace (\mathbf{n}_p) $. If we now evolve both $\mathbf{n}_q$ and $\mathbf{n}_p$ under the \emph{same} $U_{23}$ transformation to get $\mathbf{n}_p(t)$ and $\mathbf{n}_q(t)$ then their traces will remain identical. That is,
\begin{equation}\trace ( \mathbf{n}_q(t) )  = \trace ( \mathbf{n}_p(t) )\label{lippl} \end{equation}
\noindent Let us therefore look at the part of the expression (\ref{pin}) that varies between $p$ and $q$:
\begin{equation} \trace ( \ketbra{0}{0} ( \mathbf{b}_i \mathbf{n}_j )) = \av{\mathbf{b}_i \mathbf{n}_j} \label{hooo} \end{equation}
\noindent Now let us consider the case when $j=p$. If $p$ is such that (\ref{hooo}) is zero then that part of the overall sum is zero and we do not need to consider it. If $p$ is such that (\ref{hooo}) is nonzero, then let us define $\mathbf{b}_i = \mathbf{B}_i .\mathbf{\sigma} ,  \mathbf{n}_p = \mathbf{N}_p . \mathbf{\sigma}$. Then
$$ \av{\mathbf{b}_i \mathbf{n}_p}  =  \av{\mathbf{B}_i .\mathbf{\sigma} \ \mathbf{N}_p . \mathbf{\sigma}} =  \mathbf{B}_i . \mathbf{N}_p$$
\noindent Remembering that $i$ is fixed, we define the vector $ \mathbf{N}_q$ as the rotation by $\pi$ of $ \mathbf{N}_p$ about $\mathbf{B}_i$. This gives us $ \mathbf{B}_i . \mathbf{N}_q = - \mathbf{B}_i . \mathbf{N}_p$. In other words, for any given $j=p$ we can always find $j=q$ such that $ \av{\mathbf{b}_i \mathbf{n}_q} = - \av{\mathbf{b}_i \mathbf{n}_p}$. Putting this together with the results (\ref{mptq}), (\ref{tothmptq}) and (\ref{lippl}) we see that, as promised, (\ref{pfi}) is correct.
We can now conclude that for every component of the sum (\ref{pin}) which satisfies (\ref{mel}) there is another component which cancels it. In other words, the elements of the sum where (\ref{mel}) is satisfied sum to zero and are therefore neglected.

Let us now turn to the case in which (\ref{mel}) is not satisfied, in other words where
\begin{equation} \trace ( \ketbra{0}{0} ( \mathbf{b}_i \mathbf{n}_j )) = \trace ( \ketbra{0}{0} \mathbf{b}_i ) \trace ( \ketbra{0}{0} \mathbf{n}_j ) \label{maa} \end{equation}
\noindent In this situation (\ref{pin}) becomes
\begin{eqnarray*} \lefteqn{\sum_j \trace ( \mathbf{m}_j \otimes \mathbf{n}_j  \ U_{23}^\dagger \mathbf{\sigma} \otimes \iden U_{23})}\\
& & {}. \trace ( \ketbra{0}{0} \mathbf{m}_j ) \trace ( \ketbra{0}{0}  \mathbf{b}_i ) \trace ( \ketbra{0}{0} \mathbf{n}_j )\end{eqnarray*}
\noindent We can neglect the unvarying element  $\trace ( \ketbra{0}{0}  \mathbf{b}_i )$:
\begin{eqnarray} \lefteqn{\sum_j \trace ( \mathbf{m}_j \otimes \mathbf{n}_j  \ U_{23}^\dagger \mathbf{\sigma} \otimes \iden U_{23}) \trace ( \ketbra{0}{0} \mathbf{m}_j )  \trace ( \ketbra{0}{0} \mathbf{n}_j )}\nonumber\\ 
& = &  \sum_j \trace ( \mathbf{m}_j \otimes \mathbf{n}_j  \ U_{23}^\dagger \mathbf{\sigma} \otimes \iden U_{23}) \trace ( \ketbra{00}{00} \mathbf{m}_j \otimes \mathbf{n}_j ) \nonumber \\
{} & = &\sum_j   \nu_j \av{[\mathbf{q}_2^j]_{23}} \label{got} \end{eqnarray}
\noindent where the normalisation is explicitly neglected in the last step. We note that each component in the sum is non-negative, so the total cannot be zero.

Now in order to combine this expression properly with the expression for varying $i$ we need to look closely at (\ref{maa}). This is not just a condition on the $j$-components of the sum, but on both the $i$ and $j$. As a consequence, we are summing not over individual $i$ and $j$ indices, but over a third index (which we will call $k$) which denotes the \emph{pairs} of $i$ and $j$ which enable (\ref{maa}) to be satisfied. We therefore replace both the $i$ and $j$ indices with $k$. We therefore have (\ref{got}) as
$$ \sum_k   \nu_k \av{[\mathbf{q}_2^k]_{23}}$$
Let us now combine this with the expression that we had for $i$ in (\ref{tins}). As we replace $i$ with $k$ we get
\begin{eqnarray*} \lefteqn{\av{\mathbf{q}_1 \mathbf{q}_2} }\\
& = & \sum_{ij} \trace ( \mathbf{a}_i \otimes \mathbf{b}_i \ U_{13}^\dagger \mathbf{\sigma} \otimes \iden U_{13})  \trace ( \mathbf{m}_j \otimes \mathbf{n}_j  \ U_{23}^\dagger \mathbf{\sigma} \otimes \iden U_{23})   \\
& & {}.\trace ( \ketbra{0}{0} \mathbf{a}_i ) \trace ( \ketbra{0}{0} \mathbf{m}_j ) \trace ( \ketbra{0}{0} ( \mathbf{b}_i \mathbf{n}_j )\\
& = & \sum_k \trace ( \mathbf{a}_k \otimes \mathbf{b}_k \ U_{13}^\dagger \mathbf{\sigma} \otimes \iden U_{13})\\
& & {}. \trace ( \ketbra{0}{0} \mathbf{a}_k ) \trace ( \ketbra{0}{0} \mathbf{b}_k )  \nu_k \av{[\mathbf{q}_2^k]_{23}} \\
& = & \sum_k \trace ( \mathbf{a}_k \otimes \mathbf{b}_k \ U_{13}^\dagger \mathbf{\sigma} \otimes \iden U_{13}) \\
&  & {}.\trace ( \ketbra{00}{00} \mathbf{a}_k \otimes \mathbf{b}_k )  \nu_k \av{[\mathbf{q}_2^k]_{23}} \\
& = & \sum_k  \mu_k \av{[\mathbf{q}_1^k]_{13}}  \nu_k \av{[\mathbf{q}_2^k]_{23}} \\
& = & \sum_k \lambda_k \av{[\mathbf{q}_1^k]_{13}} \av{[\mathbf{q}_2^k]_{23}}
\end{eqnarray*}
\noindent which is the known separation criteria for mixed states. 

We have therefore shown that if we can write
\begin{equation}\av{\mathbf{q}_1 \mathbf{q}_2} = \sum_{ij} \lambda_{ij} \trace ( \ketbra{000}{000} \ U_i^\dagger \mathbf{\sigma} \otimes \iden U_i \otimes \iden \ \ \iden \otimes U_j^{\prime \dagger} \iden \otimes \mathbf{\sigma} U_j^\prime )\label{sepcritter}\end{equation}
\noindent where
\begin{eqnarray*} \av{\mathbf{q}_1} & = & \sum_i \mu_i \trace ( \ketbra{00}{00} U_i^\dagger \mathbf{\sigma} \otimes \iden U_i ) \\
\av{ \mathbf{q}_2} & = & \sum_j \nu_j \trace ( \ketbra{00}{00} U_j^{\prime \dagger} \iden \otimes \mathbf{\sigma} U_j^\prime )
\end{eqnarray*}
\noindent then the systems 1 and 2 are separable, and if their descriptors cannot be written in this way then they are entangled. In terms of the reduced descriptors, this separability condition becomes
$$ \av{\mathbf{q}_1 \mathbf{q}_2} = \av{\sum_{ij} \lambda_{ij} [\mathbf{q}_1^i]_{13}[ \mathbf{q}_2^j]_{23} }$$
\noindent where
$$ \av{\mathbf{q}_1} = \av{\sum_i \mu_i [\mathbf{q}_1^i]_{13}}, \ \ \ \av{\mathbf{q}_2} = \av{\sum_j \nu_j [\mathbf{q}_2^j]_{23}  }$$

\section{Physical entanglement}

This separability criterion that we have derived gives us a notion of mixed state entanglement that is an extension of that which we had for pure states. Again, reduced descriptors play an key role, but there is an important difference here. Instead of reducing the space of the descriptors to that of one qubit, we here reduce them to that of two. That is because we are really dealing with \emph{three} qubits: the two which are in a mixed state, and the third which purifies them. We will see that this third qubit is not merely a mathematical convenience, but plays an important physical role in entanglement and correlation. Despite this, the esential point remains the same: if the two qubits are separable, then the description of one qubit does not depend on the description of its past interaction with the other qubit. Put another way, all the physical properties of one qubit can be described without reference to the physical properties of the other at the time at which they previously interacted. 

What this means is that if we have two qubits, 1 and 2, which are entangled, then qubit 1 contains data about qubit 2, and \emph{vice-versa}. This information is what was received during their mutual past interaction. That is, {\bfseries shared entanglement enables one qubit to carry information about another}.

It is important to compare this characterization of entanglement with correlation. Consider two mixed qubits (and a third which purifies the system), which are separable but correlated. We can represent their descriptors as in (\ref{hip}) and (\ref{hop}), and they satisfy (\ref{sepcritter}). We can therefore write the correlation condition $ \av{\mathbf{q}_1 \mathbf{q}_2} \neq  \av{\mathbf{q}_1} \av{ \mathbf{q}_2}$ as
\begin{eqnarray*} \lefteqn{\sum_{ij} \lambda_{ij} \trace ( \ketbra{000}{000} \ U_i^\dagger \mathbf{\sigma} \otimes \iden U_i \otimes \iden \ \ \iden \otimes U_j^{\prime \dagger} \iden \otimes \mathbf{\sigma} U_j^\prime )} \\
& & \neq \sum_{ij} \mu_i \nu_j \trace ( \ketbra{00}{00} U_i^\dagger \mathbf{\sigma} \otimes \iden U_i )   \trace ( \ketbra{00}{00} U_j^{\prime \dagger} \iden \otimes \mathbf{\sigma} U_j^\prime )\end{eqnarray*}

We can see that the only place the inequality can come from is the combination of descriptors on $\h{3}$. This gives us the principal difference between correlation and entanglement: correlation may always be represented as occurring through a third party, entanglement may never be. This is because we can always represent the correlation between two qubits as arising because they each, separately, have interacted with, and therefore contain a description of, a third qubit. Entanglement, however, cannot be represented in this way. In other words, if two qubits are \emph{entangled} then they each carry information about the other, but if two qubits are \emph{correlated} then they carry information not about each other, but about a third qubit.

This is, in fact, a considerable advantage of our new conception of entanglement. Standardly, entanglement and correlation are difficult to separate conceptually, and the same condition -- $\av{AB} \neq \av{A} \av{B}$ -- does double work for both pure-state entanglement and mixed-state correlation. It is difficult to say exactly in what the difference between them consists, except to describe one as quantum and the other as classical, which is somewhat circular.

The second advantage, which seems at first sight to contradict the first, is that we can see that entanglement and correlation arise from exactly the same mechanism. Being able to describe both in the same framework is a great improvement on having to think of them as entirely separate. In the understanding of entanglement that we are advancing here, correlation arises because of entanglement. For two (mixed) systems to be correlated, they both must have interacted with a third system (this purifies the overall system). Systems 1 and 2 are then both, separately, entangled with system 3. Suppose we now ignore system 3, or we do not have access to it. The separate information that system 1 and 2 carry about system 3 can then manifest itself as a correlation between 1 and 2 when they come into contact. A concrete example would be ``system 3'' as a person giving out one half of a playing card to another two people. When these two come together later, they find that their cards are correlated, because they each had separately interacted with ``system 3". 

As a consequence of this, we can also use our conception of entanglement to help with understanding the notion of \emph{classicality}. One part of a definition of the `classical domain' is that there is no access to entanglement, only correlation. We can see now in greater detail what this entails. Firstly, it is not simply the case that there is no entanglement at all -- then there would be no correlation. Rather, we can say that systems which may be described classically are those whose descriptions rely only on reference to themselves and to their own past evolution, or to systems which are no longer accessible. There is no irreducible part of the description of one system that includes another system with which it is still interacting. Systems in situations which can be described classically can be described essentially as separate -- this description relies only on their own past evolution (or on that of systems which are now inaccessible). Systems which are entangled cannot be described in this separate way. An archetypal example of a classical situation is when a system has undergone decoherence. In this situation `system 3' is the environment to which systems 1 and 2 have been entangled. If we look just at systems 1 and 2 then they behave separately from each other (they would not behave separately from system 3, but we have no access to it). They may be correlated, but not entangled. 

So far, for reasons of tractability, we have been considering two-party entanglement. However, one bonus of our approach is that even in the bipartite case we also have elements of mutipartite entanglement. This makes it a lot easier in this approach to consider the extension of an understanding of entanglement to the mutlipartite case. Let us start by considering the standard 3-qubit system that we have been using. The entanglement of three qubits may either be simply bipartite entanglement, or it may take the form of `true' tripartite entanglement \cite{trip}. There may also be a combination of the two. True tripartite entanglement comes in two distinct forms, typified by the GHZ and W states. We saw the descriptors for the GHZ state at the beginning of this paper, and we will now look at the (unnormalised) form for the W state:
\begin{eqnarray*}\mathbf{q}_1 & = & \left( \sigma_z \otimes \sigma_x \otimes \sigma_x, \  \sigma_y \otimes \sigma_x \otimes \sigma_x ,\  -\sigma_x \otimes \iden \otimes \iden \right)\\
\mathbf{q}_2 & = &  \left(  \sigma_x \otimes \sigma_z \otimes \sigma_x ,\ - \iden \otimes \sigma_y \otimes \sigma_x ,\ \sigma_x \otimes \sigma_x \otimes \iden \right)\\
\mathbf{q}_3 & = & \left(  \iden \otimes \iden \otimes \sigma_x ,\ \iden \otimes \sigma_x \otimes \sigma_y ,\ \iden \otimes \sigma_x \otimes \sigma_z  \right)          \end{eqnarray*}    

On our understanding of entanglement, we would expect true tripartite entanglement to come about when a qubit contains information not only about itself but also about two other qubits (contrast with the bipartite case where it contains data about one other qubit). We would not expect \emph{all} qubits to demonstate this property as this would pre-judge the question of whether there is bipartite entanglement as well. This is indeed what we see in both the GHZ and W states. In the GHZ case, qubit 1 contains information about both qubits 2 and 3, and in the W state case both qubits 1 and 2 contain complete information (remembering of course that the numbering of the qubits here is arbitrary; both states have been constructed using a non-unique circuit). We would expect this to be the case for larger numbers of qubits and greater orders of entanglement. A true n-partite entanglement would arise from at least one qubit containing information about all the other n-1 qubits. 

\section{Locality}

We have seen how entangled systems contain information about each other, and seen the implications of this for various aspects of quantum information theory. The problem now is, where does this leave us with the question of the locality or otherwise of entanglement? 

Entanglement is usually considered to be a nonlocal effect because of the apparently intantaneous (or at least faster-than-light) way in which a change in one system communicates itself to another with which it is entangled. For example, in a Bell-type experiment we separate a singlet pair of electrons and send them to different detectors. A choice of basis at one detector will project out the electron heading towards the other detector on timescales shorter than that of a classical signal passing between them. Such a nonlocal effect is absolutely necessary for the individual outcomes to be correlated at the different detectors. This, combined with the Bell results for the impossibility of a local hidden-variables theory \cite{bell,aspect}, appears to give the non-local nature of entanglment the status of a fact about the world. How, then, can we claim that a different \emph{formalism} will change this?

The answer to this rests on two important points. The first is that we need to take a closer look at what it really means for the outcomes of measurements on systems to be correlated in a stronger-than-classical way. Taking the Bell-type senario of the previous paragraph, if we look simply at the results of the two detectors then we are told nothing -- the no-signalling theorem \cite{nosig} requires the statistics at one not to change when settings on the other are changed. What we must do is \emph{compare} the individual measurement outcomes at the two detectors -- which requires the results to be communicated to the same place (perhaps another researcher). It is only then, in a purely \emph{local} operation, that the outcomes are found to be correlated \cite{chrisharvey}. We can think of this as a consequence of the lack of collapse during the detecting processes: a particular choice of outcome at that time is not needed, and so we do not require a superluminal signal to pass between the systems.

Although this first point is important, it shows only that we need not have nonlocality when we have entanglement; it does not show that we do not in fact have it. We can easily imagine a senario in which there is no collapse but still nonlocal signalling between systems. We now turn to our second important point, which comes directly from our analysis of descriptors. This is that entangled systems carry information about one another, and that the information that one system carries is data about the state of the other systems \emph{at a previous time}, that is, when they last interacted. Information is not carried about the \emph{current} state of the other system, and we can see clearly from the evolution of the descriptors that an operation on one system alone changes nothing about the other. What it \emph{will} change, however, is the state of the joint system when the individual systems (or information about them) come back into contact. The mutual information contained in the systems will then determine how they combine, and it is here that entanglement correlations are displayed. This is a purely local description; at no point does information flow nonlocally.

Before we continue, it is interesting to note a lemma to these main points. We have seen that it is when we build a joint state out of individual descriptors that we see the correlations of entanglement. We saw also that this situation can take the form of information about detector outcomes being transmitted to a mutual localtion. This transmission is entirely classical: for example, it could take the form of one researcher talking on the 'phone to another. We therefore see that, with our understanding of entanglement, we also have an picture of quantum information being transmitted through classical channels. The data that entangled systems contain about each other is capable of transmission through what is usually thought of as a classical channel.

This is, in fact, something that should be expected, given our previous discussion of classicality on this view. This is yet another example of the lack of a sharp distinction between `quantum' and `classical', and the rather messy, `for most current practical purposes' definition of the classical realm. It is an interesting insight into what is meant by a `bit' however -- while bits may not take part in entangling operations, they may nevertheless carry information that was gained that way. This operation of a bit is crucial to the locality of entanglement.

Let us now look at how this local view of entanglement works in practice. We will take a qubit pair in the entangled state $\ket{00} + \ket{11}$ and send them to different detectors. The detectors start in the $\ket{0}$ state and perform perfect measurements of qubits 1 and 2 in the computational basis, and then transmit this information through a classical channel to the researcher. We start with the entangled qubits:
\begin{eqnarray*}\mathbf{q}_1 & = & \left( \sigma_z \otimes \sigma_x, \ - \sigma_y \otimes \sigma_x, \ \sigma_x \otimes \iden \right)\\
\mathbf{q}_2 & = &  \left(  \iden \otimes \sigma_x,\  \sigma_x \otimes \sigma_y,\ \sigma_x \otimes \sigma_z \right)\end{eqnarray*}
\noindent As we have seen, each carries information about the other (signified here by the $\sigma_x$ components). Now we separate them, and perform perfect measurements (CNOT operations) on them, with the results in qubits 3 and 4 respectively:
\begin{eqnarray*}\mathbf{q}_3 & = & \left( \iden \otimes \iden \otimes \sigma_x \otimes \iden, \ \sigma_x \otimes \iden \otimes \iden \otimes \sigma_y, \ \sigma_x \otimes \iden \otimes \iden \otimes \sigma_z \right)\\
\mathbf{q}_4 & = &  \left( \iden \otimes \iden \otimes \iden \otimes \sigma_x, \ \sigma_x \otimes \sigma_z \otimes \iden \otimes \sigma_y, \ \sigma_x \otimes \sigma_z \otimes \iden \otimes \sigma_z \right)\end{eqnarray*}
\noindent We now send the results of the measurements down a classical channel. This is represented by using only the $q_z$ components of the descriptors:
\begin{eqnarray*}\mathbf{q}^{\mathrm{BIT}}_3 & = & \left( \sigma_x \otimes \iden \otimes \iden \otimes \sigma_z \right)\\
\mathbf{q}^{\mathrm{BIT}}_4 & = &  \left(  \sigma_x \otimes \sigma_z \otimes \iden \otimes \sigma_z \right)\end{eqnarray*}
\noindent We can see from these the information about qubit 1 travelling down this classical channel. It is how these dependencies (one straight from qubit 1, and one that has come via qubit 2) combine that shows the correlations due to entanglement in the measurement results. This process, and the transmission of information about the past states of the entangled qubits, is all entirely local.

\section{Conclusion}

In this paper we have seen how the descriptor formalism gives us a view of entanglement where two systems that are entangled carry information about each other. We have seen how this differs significantly from correlation, in which situation the systems would each carry information about another, third, system. We have also seen how this means that correlation arises from entanglement, and the implication for a notion of classicality. We have explored how this gives insight into the relation between subsystems and entangled larger systems, and the information about different systems that is carried. We have expanded this understanding to the multipartite case. Finally, we have seen how this understanding of entanglement is all entirely local.

This is a much more comprehensive picture of entanglement than has hitherto been given.  It is a coherent conceptual picture of what is usually thought of as a ``mystery" of quantum mechanics. The use of a Heisenberg picture aided our understanding because it did not contain the artificial non-locality of the Schr\"{o}dinger formalism, and so we were able to gain a conception of entanglement from a Heisenberg separability criterion that we would not be able to find from any of the standard criteria. The discovery that it is possible to do without nonlocality in characterizing entanglement is important for our understanding both of entanglement and also of the nature of what is usually described as nonlocality in quantum mechanics. 

This novel understanding of entanglement has implications for a very wide ranger of areas, of which we will mention just a few. In the foundations of quantum mechanics it can help with understanding the quantum/classical divide, and with exploring important entanglement-based situations (such as the Bell inequality set-ups). In quantum computation, it gives us a much firmer understanding of a vital resource, and suggests that we may be able to use the idea of a `flow' of dependencies to show which qubits are in use during any given part of an algorithm or protocol. We have also shown that the resources of entanglement and nonlocality are different, and situations where they are used separately may point us towards new information transmission protocols. As this understanding has also given us a greater conception of what is meant by a classical computation, it points towards a greater knowledge of the fundamental nature of \emph{quantum} computation and how far it may go beyond what is possible classically.

\end{document}